\documentclass[12pt]{article}
\usepackage{times}

\usepackage[dvips]{graphicx}
\usepackage[usenames,dvipsnames]{color}
\usepackage{amssymb}
\usepackage{bez123}
\usepackage{amssymb}
\usepackage{floatflt}
\usepackage{wrapfig}
\usepackage{listings}
\usepackage{multirow}

\usepackage[all]{xy}

\usepackage[T1]{fontenc}
\usepackage{calc}
\usepackage{setspace}
\usepackage{color}

\begin{document}

\title{Journal Status}

\author{ $\mbox{Johan Bollen}^{\star\dagger}\mbox{, Marko A. Rodriguez}^\dagger \mbox{, and Herbert Van de Sompel}^\dagger$ }
\maketitle

\vspace{-1cm}
\begin{center}
        {\small {\it $^\dagger$ Digital Library Research \& Prototyping Team, \\Research Library, Los Alamos National Laboratory, Los Alamos, NM, 87545}}\\
	{\small $\star$ Corresponding author. Tel.: +1 505 606 0030. URL: http://public.lanl.gov/jbollen,\\
	 email: $\{$jbollen, marko, herbertv$\}$@lanl.gov}
\end{center}

\abstract{
	The status of an actor in a social context is commonly defined in terms of two factors: the total number of endorsements the actor receives from other actors and the prestige of the endorsing actors. These two factors indicate the distinction between popularity and expert appreciation of the actor, respectively.  We refer to the former as popularity and to the latter as prestige.  These notions of popularity and prestige also apply to the domain of scholarly assessment.  The ISI Impact Factor (ISI IF) is defined as the mean number of citations a journal receives over a 2 year period.  By merely counting the amount of citations and disregarding the prestige of the citing journals, the ISI IF is a metric of popularity, not of prestige.  We demonstrate how a weighted version of the popular PageRank algorithm can be used to obtain a metric that reflects prestige.  We contrast the rankings of journals according to their ISI IF and their weighted PageRank, and we provide an analysis that reveals both significant overlaps and differences.  Furthermore, we introduce the Y-factor which is a simple combination of both the ISI IF and the weighted PageRank, and find that the resulting journal rankings correspond well to a general understanding of journal status.
}

 \thispagestyle{empty}
  
%\tableofcontents
\newpage

\section{Introduction}

Some people are popular but not prestigious and vice versa. For example, an author of pulp detectives may sell many books, but may not have earned the respect of literary critics. Conversely, a Nobel Prize in Literature winner may be highly valued among literary experts, yet never make the New York Times bestseller list.  In essence, these examples reveal the existence of two factors that contribute to the status of an actor in a social context: the total number of endorsements the actor receives from other actors, and the prestige of the endorsing actors.  In the remainder of this paper, we refer to the former as popularity and to the latter as prestige.\\

Similar considerations apply to the assessment of scholarly communication where citation counts are commonly used as an indication of scholarly status. For example, a journal that publishes mostly review articles may be frequently cited by graduate students, yet largely be ignored by experts interested in the cutting edge of research. The Thomson ISI Impact Factor (ISI IF) is generally accepted as an indicator of journal status, and is defined as the mean number of citations to articles published in a journal over a 2 year period \cite{citati:garfield1979,impactreivew:garfield1999}.  Given that the ISI IF is based on the amount of citations to a journal, and does not take into account the prestige of the citing journals, it seems to only represent the popularity factor of status, not its prestige factor.\\

Many concerns have been expressed over the usefulness of the ISI IF as an indicator of journal status \cite{impact:seglen1997,impact:harter1997,assess:nederhof2001,resear:grant2002,impact:saha2003,notsodeep:nature}.  In fact, its focus on popularity would render it impossible to use in many other areas, such as for example the WWW. A web page that is often linked to can indeed be of very low status and vice versa. For that reason, alternatives to link counting, developed in the domain of social network analysis, have been widely adopted for WWW searching.\\

When the Google search engine ranks web pages according to their status it does so by not merely counting the number of hyperlinks to a page. Google's PageRank algorithm \cite{anatom:brin1998} computes the status of a Web page based on a combination of the number of hyperlinks that point to the page and the status of the pages that the hyperlinks originate from.  By taking into account both the popularity and the prestige factor of status, Google has been able to avoid assigning high ranks to popular but otherwise irrelevant Web pages.\\

The success of the PageRank algorithm in the Google environment has led to PageRank becoming a standard technique to assess the status of web resources. However, where the evaluation of journal status is concerned the ISI IF still rules supreme. This situation may not be sustainable. As an ever growing collection of scholarly materials becomes available on the Web, and hence becomes searchable through Google and Google Scholar, our perception of article status (and hence of journal status) will change as a result of the PageRank-driven manner by which Google lists its search results.  In the future, PageRank, not the ISI IF, may very well start representing our perception of article and journal status.\\

A change from the ISI IF to PageRank-based metrics for journal ranking would effectively signify a shift from an evaluation based on popularity, i.e.~citation frequency, to an evaluation based on prestige, i.e.~the prestige of those who cite is taken into account. To evaluate the consequences of such a change on the assessment of journal status, we used the dataset of the 2003 ISI Journal Citation Reports (ISI JCR) to compare the ISI IF and Weighted PageRank rankings of journals.  We paid special attention to journals with a significant discrepancy between their ISI IF and Weighted PageRank values. We also introduce a ranking principle, the Y-factor, to rank journals according to whether they have both high ISI IF and Weighted PageRank values.

\section{Two common metrics of status}

Citations are at the basis of most present attempts to assess scholarly impact.  This is true for the assessment of the impact of individual articles, journals, researchers \cite{itali:calza1995,index:ball2005,indexq:hirsch2005}, research departments, universities and even countries \cite{indica:braun1996,advant:bordons2002,resear:rey2002,journa:kaltenborn2003}.  As articles cite one another, they define an article citation network in which each node represents an article and each directed edge represents a citation by that article to another.  By grouping all articles published in the same journal under a single journal node, an article citation network can easily be transformed into a Journal Citation Network.  In that network, the directed edges between the journal nodes represent the collection of citations from one journal to another. This network can be formalized as a set of journals $V$, a set of directed edges $E \subseteq V^2$ that exist between the journals in $V$, and the function $W(v_i,v_j) \rightarrow \mathbb{N}^+$ which maps each edge between the journal $v_i$ and $v_j$ to a positive, integer citation frequency. A range of journal status metrics can be applied to such a Journal Citation Network. In the following sections, we discuss two highly common metrics, namely the ISI Impact Factor and Google's PageRank. The latter has been modified to take into account edge weights (Weighted PageRank) so that it can be applied to the Journal Citation Network.

\subsection{The ISI Impact Factor}

The ISI IF defines the status of a journal for a specific year as the mean number of citations that occurred in that year to the articles published in the journal during the two previous years. More concretely, the 2003 ISI IF of a journal $v_i$ is calculated by dividing the number of citations made in 2003  to $v_i$'s 2001 and 2002 articles by the total number of articles $v_i$ published in 2002 and 2001.  Expressed in terms of a Journal Citation Network the ISI IF corresponds to a journal's in-degree \cite{indegree:alexander} normalized by the total number of papers the journal published in that period. Eq. \ref{IF} defines the IF of journal $v_i$ in year $t$, labeled $IF(v_i,t)$, as follows:

\begin{equation}
IF(v_i,t) = \frac{\sum_j c(v_j,v_i,t)}{n(v_i)}
\label{IF}
\end{equation}

where $c(v_j, v_i, t)$ corresponds to the number of citations from journal $v_j$ to journal $v_i$ in year $t$. The number of publications published in journal $v_i$, denoted $n(v_i)$, during the two years previous to $t$, normalizes the resulting citation count, leading to a mean, 2-year citation rate per article.\\

In social network analysis terms, in-degree can be considered a metric of popularity because it corresponds to the number of endorsements received by a particular actor in the network.  And, indeed, when assuming that a citation to a journal indicates an endorsement of the journal's content, we find that, in terms of social network analysis, the ISI IF is a measure of popularity because a journal has a higher ISI IF if its articles are more often cited.

\subsection{Journal PageRank}

This aspect of the ISI IF has been known and studied for decades. In particular, Pinski et al. (1977) \cite{citati:pinski1976} propose an algorithm that evaluates the influence of journals by taking into account not simply the number of citations from one journal to the other, but also the prestige of the citing journal. Journals that receive many citations from prestigious journals are considered highly prestigious themselves. By iteratively passing prestige from one journal to the other, a stable solution is reached which reflects the relative prestige of journals. This procedure is highly related to efforts in social network science to define status in terms of "inherited" status, e.g. eigenvector centrality \cite{power:bonacich1987} , systems to separate web pages into "authoraties" and "hubs" \cite{author:kleinberg1998,hubs:kleinberg1999} and recent investigations of the role of journals as knowledge sources or storers \cite{assess:nerur2005}.\\

Within this long lineage of social network metrics of status, the founders of the Google search engine outline an algorithm to assess the prestige of web pages based on similar principles in their 1998 paper "The anatomy of a large-scale search engine" \cite{anatom:brin1998}. This concept is further developed in later publications \cite{pagera:page98} in terms of random walk models of web navigation. Much like the proposal by Pinski et al., PageRank is calculated by an iterative algorithm which propagates prestige values from one web page to another and converges to a solution \cite{perron:pillai2005}. The PageRank equation that governs the iterative transfer of PageRank values from one web page to the other is shown in Eq. \ref{PReqb}:

\begin{equation}
PR(v_i)= \frac{(1-\lambda)}{N} + \lambda \sum_j PR(v_j) \times \frac{1}{O(v_j)}
\label{PReqb}
\end{equation}

In Eq. \ref{PReqb}, it is assumed that a collection of pages $v_j$ link to a recipient page $v_i$ and each transfers a proportion of their PageRank, denoted $PR(v_j)$, to $v_i$. It is also assumed that PageRank values are equally distributed along a page's out-links, i.e.~if a page $v_j$ has 3 out-links each recipient page $v_i$ receives only one-third of $v_j$'s PageRank. Transfered PageRank values are therefore normalized by the number of out-links from page $v_j$ which is denoted $O(v_j)$. The parameter $\lambda$, which can take values between zero and one, represents the attenuation of prestige values as they are transferred from one web page to the other. The parameter $\frac{(1-\lambda)}{N}$ represents the minimal amount of prestige assigned to each web page.  $N$ represents the total number of pages in the network.

%Following Eq. \ref{PReqb}, random prestige values can initially be set for all web pages in a collection.  Then, several iterations of Eq. \ref{PReqb} can be applied, and, after a finite number of iterations, the computed PageRank values will converge to their final values. This convergence is mathematically guaranteed by the Perron-Frobenius theorem \cite{perron:pillai2005}.\\

\subsection{Weighted PageRank for Journal Citation Networks}

PageRank has become a standard to evaluate the status of web pages. It is our objective to apply it to Journal Citation Networks so that we can compare two highly common metric of status, i.e.~the ISI IF and PageRank, in terms of their ability to evaluate the relative popularity or prestige of journals. The PageRank definition above assumes that prestige is distributed equally across all of a web page's hyperlinks. This is appropriate since hyperlinks are not weighted, i.e. each hyperlink indicates an equal degree of relationship between a pair of linked pages. In the Journal Citation Network, however, not all edges are created equal; some journals are connected by more citations than others. The PageRank equation when applied to journal citation networks should therefore be adapted to take into account journal citation frequencies in its transfer of PageRank values.  Indeed, if a journal $v_j$ cites journal $v_i$ 10 times more frequently than any other journal, the amount of prestige transferred from $v_j$ to $v_i$ should be ten times as high.  More generally, a journal that receives many citations from a specific other journal should receive a matching proportion of that journal's prestige.\\

This is in fact a common problem encountered in applications of PageRank to weighted networks. Modifications of the PageRank equation have therefore been proposed to take into account link weights. A Web-based Weighted PageRank algorithm has earlier been defined \cite{prw:xing2004} to calculate aggregate web site prestige and a weighted PageRank algorithm has been to used to rank authors in a weighted co-authorship network \cite{coauth:liu2005}. The notion of weighted link weights is in fact an integral part of Pinski and Narin's approach to define journal prestige. We will briefly discuss these common modifications of the PageRank equation in terms of weighted journal citation networks below.\\

Assume we need to rewrite Eq. \ref{PReqb} to account for the transmission of journal prestige relative to the number of citations that exist between pairs of journals in the Journal Citation Network.  First, we define a propagation proportion $w(v_j,v_i)$ between journals $v_i$ and $v_j$ by normalizing the link weights emanating from a particular journal $v_j$ as follows:

\begin{equation}
w(v_j,v_i) = \frac{W(v_i,v_j)}{\sum_k W(v_j,v_k)}
\label{citnorm}
\end{equation}.

For any particular journal $v_j$, all $w(v_i,v_j)$ now sum up to one and it can therefore be used to determine the fraction of a journal's PageRank it transfers to the journals it cites.\\

We now obtain the Weighted PageRank equation for journal $v_j$ as follows:

\begin{equation}
PR_w(v_i) = \frac{\left(1-\lambda\right)}{N} + \lambda \sum_j PR_w(v_j) \times w(v_j,v_i)
\label{PRweq}
\end{equation}

According to Eq. \ref{PRweq}, the transfer of prestige from one journal to the other is modulated by the propagation proportion $w(v_j,v_i)$.
In effect, the equal distribution of PageRank values in Eq. \ref{PReqb}, as given by the factor $\frac{1}{O(v_i)}$, has been replaced by the propagation proportion $w(v_j,v_i)$ thereby allowing Weighted PageRank to be calculated for Journal Citation Networks.

\subsection{Product of ISI IF and Weighted PageRank}

We now have two different, but highly common, metrics of status at our disposal. The ISI IF relies on citation frequencies and therefore stresses the popularity aspect of journal status. The Weighted PageRank, as defined above, relies on a propagation of prestige values from one journal to the other, and therefore corresponds better to our intuitive notion that prestige is not only a matter of the number of endorsements, but who is actually endorsing. \\

Thus defined, the ISI IF and the Weighted PR represent highly common, but possibly different, facets of journal status. As will be demonstrated in section \ref{outliers}, there can indeed exist significant discrepancies between a journal's ISI IF and Weighted PageRank values, i.e.~some journals can have high ISI IF and low Weighted PageRank values, and vice versa. To rank journals on the basis of both metrics combined we defined a product of the ISI IF and the Weighted PageRank, labeled $Y$-factor, as shown in Eq. \ref{combmetric} below.

\begin{equation}
Y(v_j) = \mbox{ISI IF}(v_j) \times \mbox{PR}_w(v_j)
\label{combmetric}
\end{equation}

Journals that score highly on the Y-factor will be ranked highly by either or both the ISI IF and Weighted PageRank. The resulting rankings are included in the following section for informational purposes.

\section{Indicators of Journal Status}

In this section, we compare three indicators of status in the Journal Citation Network: the popularity-oriented ISI IF, the prestige-oriented Weighted PageRank and a product of both, namely the Y-factor.  We do so based on the dataset provided by the 2003 ISI Journal Citation Reports.

\subsection{Comparing the ISI IF and the Weighted PageRank}

In order to obtain Weighted PageRank values for journals that have an ISI Impact Factor, a Journal Citation Network was constructed on the basis of the 2003 ISI JCR data set which contains 2003 journal citations to 2001 and 2002 publications.  This journal citation information was represented as a matrix in which both rows and columns represent journals, and in which cells represent the amount of times a journal in a row cites a journal in a column.  Not surprisingly, a sparse matrix resulted, with 5710 journals having non-zero citation counts.\\

To provide an indication of the overall characteristics of the ISI IF and the Weighted PageRank, Table \ref{ranks} shows the ten highest ranking journals for both status metrics.  Clearly, the rankings diverge significantly, with only three journals, Nature, Science and The New England Journal of Medicine being represented in both lists.  We observe that the journals with the highest ISI IF are strongly positioned in the area of medicine, with review journals being heavily represented.  The latter confirms the characterization of the ISI IF as a popularity-oriented metric, since review journals typically publish background material that is likely to be cited frequently. Overall, the listing according to Weighted PageRank shows more variations in scholarly discipline, and many of the top-ranked journals such as Science, Nature, Cell and the Journal of Biological Chemistry are generally considered highly prestigious journals.\\

\begin{table}[p]
\begin{footnotesize}
\begin{center}
\begin{tabular}{r||rr||rr||rr}
	 \multicolumn{3}{c}{ISI IF}				&		\multicolumn{2}{c}{$\mbox{PR}_w$} & 	\multicolumn{2}{c}{Y-factor}			\\\hline
rank & value	& Journal					& value (x $10^3$)	& Journal				& value(x $10^2$) 	& Journal				\\\hline
1	& 52.28	& ANNU REV IMMUNOL		& 16.78		& NATURE 				& 51.97		& NATURE				\\
2	& 37.65	& ANNU REV BIOCHEM 		& 16.39		&  J BIOL CHEM			& 48.78		& SCIENCE				\\
3	& 36.83	& PHYSIOL REV			& 16.38		& SCIENCE				& 19.84		& NEW ENGL J MED		\\
4	& 35.04	& NAT REV MOL CELL BIO	& 14.49		& PNAS				 	& 15.34 		& CELL					\\
5	& 34.83	& NEW ENGL J MED		& 8.41		&  PHYS REV LETT			& 14.88 		& PNAS					\\
6	& 30.98	& NATURE				& 5.76		& CELL 					& 10.62		& J BIOL CHEM			\\
7	& 30.55	& NAT MED				& 5.70		& NEW ENGL J MED 		& 8.49		& JAMA 					\\
8	& 29.78	& SCIENCE				& 4.67		& J AM CHEM SOC			& 7.78		&  LANCET				\\
9	& 28.18	& NAT IMMUNOL			& 4.46		& J IMMUNOL				& 7.56		& NAT GENET				\\
10	& 28.17	& REV MOD PHYS 			& 4.28		& APPL PHYS LETT			& 6.53		& NAT MED				\\\hline
\end{tabular}
\end{center}
\end{footnotesize}
\caption{\label{ranks} The highest ranking journals according to ISI IF, Weighted PageRank and Y-factor}
\end{table}

\begin{figure}[p]
\begin{center}
\includegraphics[width=6.5in]{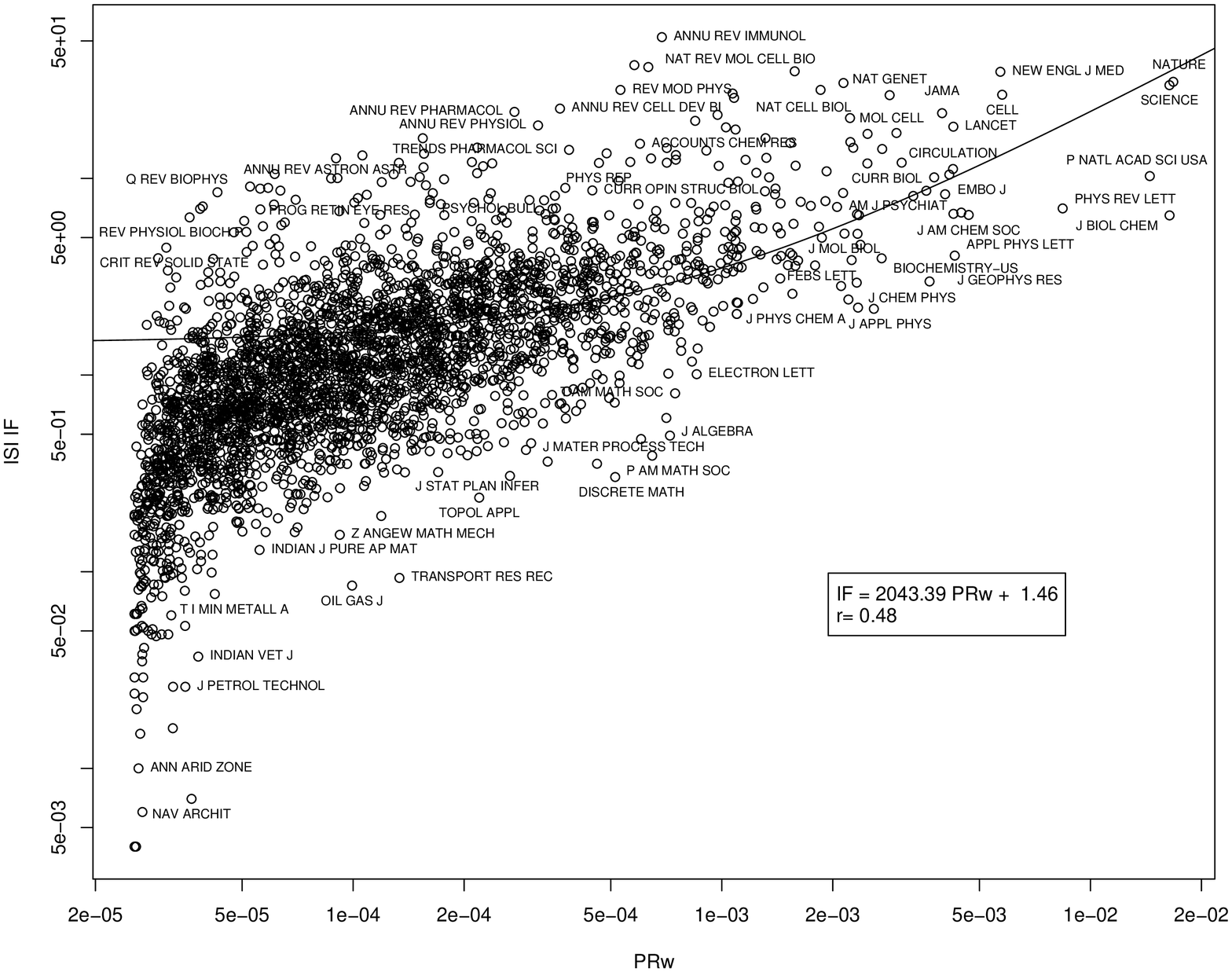}
\caption{\label{jcr03_prwif} Scatter plot of the ISI IF versus the Weighted PageRank (PRw).}
\end{center}
\end{figure}

A more quantitative analysis of the overlap and discrepancies between the two status metrics is provided by the scatter plot of Fig. \ref{jcr03_prwif}.  Despite the strong discrepancies in the top ten listings according to both status metrics, the plot reveals a significant overall correlation, confirmed by a Pearson correlation coefficient of $r = 0.48, p<0.01$ between the ISI IF and the Weighted PageRank.  We note that the journals Nature and Science are positioned in the top-right corner of the scatter plot, reflecting the fact that they have both high ISI IF and high Weighted PageRank values.  This means that both journals are often cited and are cited by prestigious journals. \\

Since it is common knowledge in bibliometrics that comparisons of ISI IF values across scholarly disciplines are problematic due to differences in the publication and citation process, we decided to focus on a subset of our Journal Citation Network that pertains to Physics journals.  We selected Physics journals in the 2003 ISI JCR dataset on the basis of the ISI subject categories listed in Table \ref{catcodes}. The resulting Physics subset of the Journal Citation Network contained 229 journals. A ranking of this subset according to ISI IF, Weighted PageRank and the Y-factor is shown in Table \ref{physicscombo}.\\

\begin{table}[p]
\begin{footnotesize}
\begin{center}
\begin{tabular}{ll}
\hline
ISI Category	& ISI Category Name \\\hline
UB     & PHYSICS, APPLIED \\
UF     & PHYSICS, FLUIDS \& PLASMAS \\
UH     & PHYSICS, ATOMIC, MOLECULAR \& CHEMICAL \\
UI     & PHYSICS, MULTIDISCIPLINARY \\
UK     & PHYSICS, CONDENSED MATTER \\
UN     & PHYSICS, NUCLEAR \\
UP     & PHYSICS, PARTICLES \& FIELDS \\
UR     & PHYSICS, MATHEMATICAL\\\hline
\end{tabular}
\end{center}
\end{footnotesize}
\caption{\label{catcodes} ISI Subject Categories for Physics Journals.}
\end{table}

\begin{table}[p]
\begin{footnotesize}
\begin{center}
\begin{tabular}{ll}
\hline
ISI Category	& ISI Category Name \\
\hline
EP	&	COMPUTER SCIENCE, ARTIFICIAL INTELLIGENCE						\\
ER	&	COMPUTER SCIENCE, CYBERNETICS									\\
ES	&	COMPUTER SCIENCE, HARDWARE \& ARCHITECTURE					\\
ET	&	COMPUTER SCIENCE, INFORMATION SYSTEMS							\\
EV	&	COMPUTER SCIENCE, INTERDISCIPLINARY APPLICATIONS				\\
EW	&	COMPUTER SCIENCE, SOFTWARE, GRAPHICS, PROGRAMMING			\\
EX	&	COMPUTER SCIENCE, THEORY \& METHODS							\\
\end{tabular}
\end{center}
\end{footnotesize}
\caption{\label{catcodesCS} ISI Subject Categories for Computer Science Journals.}
\end{table}

\begin{table}[p]
\begin{footnotesize}
\begin{center}
\begin{tabular}{ll}
\hline
ISI Category	& ISI Category Name \\
\hline
DS	&	CRITICAL CARE MEDICINE							\\
FF	&	EMERGENCY MEDICINE								\\
FY	&	DENTISTRY, ORAL SURGERY \& MEDICINE				\\
OI	&	INTEGRATIVE \& COMPLEMENTARY MEDICINE			\\
OP	&	MEDICINE, LEGAL									\\
PY	&	MEDICINE, GENERAL \& INTERNAL					\\
QA	&	MEDICINE, RESEARCH \& EXPERIMENTAL				\\
VY	&	RADIOLOGY, NUCLEAR MEDICINE \& MEDICAL IMAGING	\\
YU	&	TROPICAL MEDICINE
\end{tabular}
\end{center}
\end{footnotesize}
\caption{\label{catcodesmed} ISI Subject Categories for Medicine Journals.}
\end{table}

\begin{table}[p]
\begin{footnotesize}
\begin{center}
\begin{tabular}{c||ll||ll||ll}
\hline\hline
rank & IF	& Title						& $\mbox{PR}_w \times 10^3$	& Title			& $Y \times 10^2$	& Title		\\\hline
1	& 28.17	& REV MOD PHYS 			& 8.41	& PHYS REV LETT 				& 5.91 	& PHYS REV LETT 			\\
2	& 13.09	& ADV PHYS 				& 4.28	& APPL PHYS LETT 			&1.73	& APPL PHYS LETT 		\\
3	& 11.98	& PHYS REP 				& 2.59	& J APPL PHYS 				& 1.50	& REV MOD PHYS 			\\
4	& 10.03	& MAT SCI ENG R 			& 2.38	& PHYS REV D 				& 1.09	& PHYS REV D 			\\
5	& 8.67	& ANNU REV NUCL PART S 	& 2.34	& PHYS REV E 				& 0.69	& J CHEM PHYS 			\\
6	& 8.41	& REP PROG PHYS			& 2.32	& J CHEM PHYS 				& 0.66	& J HIGH ENERGY PHYS 	\\
7	& 7.04	& PHYS REV LETT			& 1.56	& PHYS LETT B 				& 0.63	& PHYS LETT B 			\\
8 	& 7.00	& SOLID STATE PHYS		& 1.55	& PHYS REV A 				& 0.57	& NUCL PHYS B 			\\
9 	& 6.06	& J HIGH ENERGY PHYS		& 1.22	& CHEM PHYS LETT 			& 0.56	& J APPL PHYS 			\\
10 	& 5.97	& PROG NUCL MAG RES SP	& 1.09	& J HIGH ENERGY PHYS 		& 0.56	& PHYS REP 				\\
\hline
\end{tabular}
\end{center}
\end{footnotesize}
\caption{\label{physicscombo} The highest ranking Physics journals according to ISI IF, Weighted PageRank (PRw) and Y-factor.}
\end{table}

We can detect a pattern similar to that found for the complete Journal Citation Network. Again, only 2 journals, namely Physical Review Letters and Journal of High Energy Physics, are amongst the highest ranking according to both the ISI IF and the Weighted PageRank. In addition, the ISI IF rankings can again be characterized by a preponderance of applied physics journals that frequently publish background material that is likely to be cited. The Weighted PageRank ranking seems to focus on a set of journals typically appreciated by domain experts, as the journals of the American Physical Society: Physical Review A, D and E . The Pearson correlation coefficient between ISI IF and Weighted PageRank values was found to be lower than was the case with the complete Journal Citation Network, namely $r = 0.24, p<0.01$.  This lower correlation indicates a lesser degree of intra-discipline overlap between both metrics. We leave the interpretation of the Y-factor rankings to the reader.\\

Proceeding along the same lines, we also compared the ISI IF and the Weighted PageRank for journals in Computer Science and Medicine.  Again, subsets of the Journal Citation Network were extracted by means of ISI category codes; Table \ref{catcodesCS} and Table \ref{catcodesmed} list these codes for Computer Science and Medicine, respectively. The results for Computer Science reveal an even greater discrepancy between the ISI IF and the Weighted PageRank. Indeed, as can be seen in Table \ref{ranksCS}, only the journal Bioinformatics ranks in the top ten according to both the ISI IF and the Weighted PageRank.  Again, it seems that many of the top-ranking journals according to the Weighted PageRank specialize in a focused research area. The scatterplot in Fig. \ref{CS-outliers} further confirms the greater divergence between the ISI IF and the Weighted PageRank values. The Pearson correlation coefficient was found to be $r = 0.5, p<0.01$.  The Medicine subset of the Journal Citation Network follows a different pattern than that of the Physics and Computer Science subsets.  As can be seen in Table \ref{ranksmed}, 9 journals appear in the top ten according to both the ISI IF and the Weighted PageRank.  And, the scatterplot in Fig. \ref{med-outliers} further confirms the higher degree of overlap between the two metrics in Medicine. Indeed, the Pearson correlation coefficient between the ISI IF and the Weighted PageRank values was found to be $r = 0.91, p<0.01$, indicating that the notions of prestige and popularity are more strongly intertwined for Medicine than they are for the other explored domains.  Overall, it seems that the level of discrepancy between the ISI IF and the Weighted PageRank across disciplines relates to variations in the characteristics of the publication and citation practices in different domains.

\begin{table}[p]
\begin{footnotesize}
\begin{center}
\begin{tabular}{r||rr||rr||rr}
%	 \multicolumn{3}{c}{ISI IF}				&	\multicolumn{2}{c}{$\mbox{PR}_w$} 		& 	\multicolumn{2}{c}{Y-factor}			\\\hline
\hline\hline
rank & IF	& Title						& $\mbox{PR}_w \times 10^4$	& Title			& $Y \times 10^4$	& Title				\\\hline
1 	& 7.50	& ACM COMPUT SURV		& 10.08 	& IEEE T INFORM THEORY 			& 62.27	& BIOINFORMATICS			 \\
2	& 6.70	& BIOINFORMATICS		& 9.29	& BIOINFORMATICS 				& 22.64	& IEEE T INFORM THEORY		 \\
3	& 4.54	&  VLDB J					& 5.90	& COMPUT METHOD APPL M 			& 21.68	& IEEE T PATTERN ANAL 		\\
4	& 3.87	& IEEE NETWORK			& 5.67	&  IEEE T PATTERN ANAL 			& 12.11	& ACM COMPUT SURV			\\
5	& 3.82	& IEEE T PATTERN ANAL	& 5.48	& J COMPUT PHYS 					& 11.78	& IEEE T IMAGE PROCESS		\\
6	& 3.76	& IEEE T MED IMAGING		& 4.98	& COMMUN ACM					& 11.47	& IEEE T MED IMAGING			\\
7	& 3.73	& IEEE INTELL SYST APP	& 4.95	& THEOR COMPUT SCI				& 10.37	& J ACM						\\
8	& 3.61	& IBM J RES DEV			& 4.46	& IEEE T IMAGE PROCESS			& 9.65	& J COMPUT PHYS				 \\
9	& 3.33	& INFORM SYST			& 4.35	& COMPUTER						& 8.59	& IEEE INTELL SYST APP		\\
10	& 3.32	& J ACM					& 3.36	& IEEE T NEURAL NETWOR			& 8.20	& ARTIF INTEL					 \\
\hline
\end{tabular}
\caption{\label{ranksCS} The highest ranking Computer Science journals according to ISI IF, Weighted PageRank (PRw) and Y-factor.}
\end{center}
\end{footnotesize}
\end{table}

\begin{table}[p]
\begin{footnotesize}
\begin{center}
\begin{tabular}{r||rr||rr||rr}
\hline\hline
rank & IF	& Title						& $\mbox{PR}_w \times 10^3$	& Title			& $Y \times 10^2$	& Title				\\\hline
1	& 34.83	& NEW ENGL J MED		& 5.70 				& NEW ENGL J MED			& 19.84		& NEW ENGL J MED 			\\
2	& 30.55	& NAT MED				& 4.25				& LANCET					& 8.49		& JAMA						\\
3	& 21.46	& JAMA					& 3.96				& JAMA				 		& 7.78		& LANCET 					\\
4	& 18.32	& LANCET				& 2.27				& J CLIN INVEST				& 6.53		&  NAT MED 					\\
5	& 15.30	& J EXP MED				& 2.23				& J EXP MED					& 3.42		& J EXP MED 					\\
6	& 14.31	& J CLIN INVEST			& 2.14				& NAT MED					& 3.25		& J CLIN INVEST 				\\
7	& 12.42	& ANN INTERN MED		& 1.40				& AM J RESP CRIT CARE 		& 1.44		& ANN INTERN MED 			\\
8	& 11.38	& ANNU REV MED			& 1.16				& ANN INTERN MED			& 1.24		& AM J RESP CRIT CARE 		\\
9	& 8.88 	& AM J RESP CRIT CARE	& 0.91				& NEUROIMAGE				& 0.59		& ARCH INTERN MED 			\\
10	& 6.76	& ARCH INTERN MED		& 0.87				& ARCH INTERN MED			& 0.57		& NEUROIMAGE 				\\\hline
\end{tabular}
\caption{\label{ranksmed} The highest ranking Medicine journals according to ISI IF, Weighted PageRank (PRw) and Y-factor.}
\end{center}
\end{footnotesize}
\end{table}

\subsection{Popular and Prestigious Journals}\label{outliers}

Intrigued by the significant correlation between the ISI IF and the Weighted PageRank as shown in Fig. \ref{jcr03_prwif} and the significant discrepancies revealed in Table \ref{ranks} and Table \ref{physicscombo}, we set out to inspect the Journal Citation Network for journals that have strongly diverging ISI IF and Weighted PageRank values.   Two types of divergences were explored: 

\begin{description}
	\item[Popular Journals] are journals that are cited frequently by journals with little prestige. These journals have a very high ISI IF and a very low Weighted PageRank.
	\item[Prestigious Journals] are journals that are not frequently cited, but their citations come from highly prestigious journals. These journals have a very low ISI IF and a very high Weighted PageRank.
\end{description}

\begin{figure}[p]
\begin{center}
\includegraphics[width=6.5in]{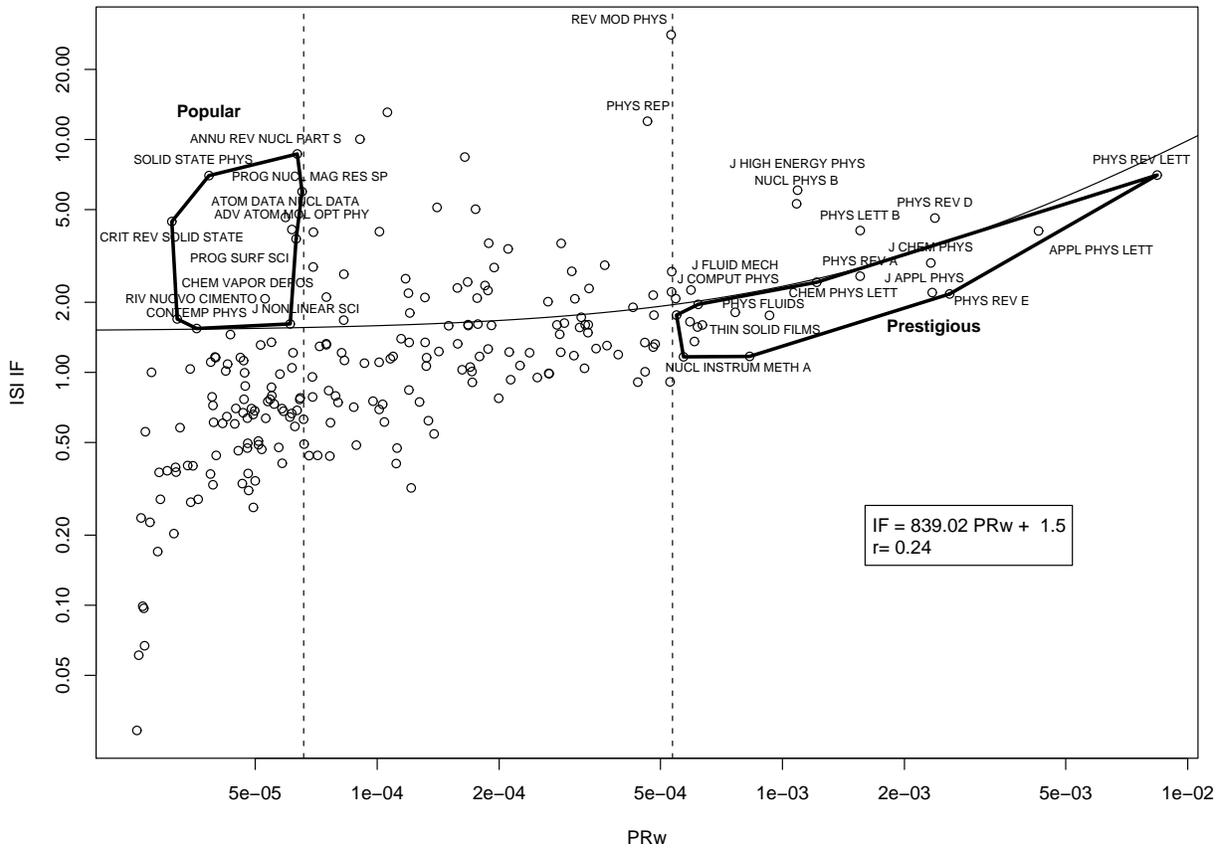}
\end{center}
\caption{\label{physics-outliers} Popular and Prestigious Journals in Physics.}
\end{figure}

We identified Popular and Prestigious Journals in the full Journal Citation Network, but were unable to recognize a meaningful pattern in the results. This was not unexpected as the exercise amounted to comparing ISI IF values across disciplines.  Hence, we decided to refocus our attention on the Physics subset of the Journal Citation Network. We empirically decided on threshold values for the ISI IF and the Weighted PageRank that guaranteed a sufficient number of journals in both the Popular and Prestigious category.  First, we decided that any Weighted PageRank value below the 40th percentile was very low, and any value above the 90th percentile was very high. These choices are represented by the vertical lines in the scatter plot of Fig. \ref{physics-outliers}. Second, to determine the low and high threshold values for the ISI IF, we generated a linear regression model for the relationship between the ISI IF and the Weighted PageRank. The result is visualized in Fig. \ref{physics-outliers} as the line that cuts across the cloud of physics journals.  Figure \ref{physics-outliers} outlines the regions of the scatter plot that correspond to our categorization of Popular and Prestigious Journals and to our chosen threshold values.  The former category is shown as the top-left region, the latter as the bottom-right region. Popular, prestigious and high Y-factor ranking journals are labeled by their abbreviated journal titles in the graph.  Table \ref{topratio} shows the ten top-ranked journals in both the Popular and Prestigious Journals category ranked by the degree  to which their actual ISI IF deviates from the value predicted by the linear regression model, labeled IF$_\Delta$.\\

\begin{table}[p]
\begin{footnotesize}
\begin{center}
\begin{tabular}{c|lccc||lccc}
	\multicolumn{5}{c}{Popular: ISI IF $\uparrow$, $\mbox{PR}_ w < 40$\%-tile}		&	\multicolumn{4}{c}{Prestigious: ISI IF $\downarrow$, $\mbox{PR}_ w > 90$\%-tile}	\\\hline
 	& Journal title				& ISI IF	& PR$_w \times 10^5$	& IF$_\Delta$				& Journal title				& ISI IF	& PR$_ w \times 10^3$	& IF$_\Delta$	\\\hline
1 	& ANNU REV NUCL PART S 	& 8.67 	& 6.35 				& 7.11 					& PHYS REV LETT			& 7.04	& 8.41 		& $-$1.52 \\
2	& SOLID STATE PHYS 		& 7.00 	& 3.85				& 5.46 					& J APPL PHYS			& 2.17	& 2.59 		& $-$1.50 \\
3	& PROG NUCL MAG RES SP 	& 5.97 	& 6.53				& 4.41 					& PHYS REV E				& 2.20	& 2.34 		& $-$1.27 \\
4	& ATOM DATA NUCL DATA 	& 4.63 	& 5.94				& 3.08					& APPL PHYS LETT			& 4.05	& 4.28 		& $-$1.05 \\
5	& CRIT REV SOLID STATE 	& 4.44 	& 3.11				& 2.91					& JPN J APPL PHYS		& 1.17	& 0.83 		& $-$1.03 \\
6	& ADV ATOM MOL OPT PHY 	& 4.11 	& 6.16				& 2.55					& NUCL INSTRUM METH A 	& 1.17 	& 0.57 		& $-$0.81 \\
7	& PROG SURF SCI 			& 3.74 	& 6.31				& 2.19					& J PHYS A$-$MATH GEN	& 1.36	& 0.61 		& $-$0.65 \\
8	& CHEM VAPOR DEPOS 		& 2.07 	& 5.29				& 0.52					& J PHYS$-$CONDENS MAT 	& 1.76	& 0.93 		& $-$0.52 \\
9	& RIV NUOVO CIMENTO 		& 1.70 	& 3.21				& 0.17					&  J CHEM PHYS			& 2.95	& 2.32 		& $-$0.50 \\
10	& J NONLINEAR SCI 		& 1.62 	& 6.10				& 0.06					& PHYS FLUIDS			& 1.57	& 0.62 		& $-$0.45 \\\hline
\end{tabular}
\end{center}
\end{footnotesize}
\caption{\label{topratio} The top-ranked Popular and Prestigious journals in Physics.}
\end{table}

A close examination of the resulting Popular Journal category reveals that it contains either review journals or journals that frequently publish data tables. Such journals are likely to be cited as background material, hence have a high ISI IF, but they do not correspond well to what would generally be perceived as a prestigious journal.  The Prestigious Journal category reveals a collection of highly esteemed Physics journals: Journal of Applied Physics, Physical Review E, and Journal of Chemical Physics to name a few.  Despite their prestigious status, they have an unexpectedly low ISI IF.\\

\begin{figure}[p]
\begin{center}
\includegraphics[width=6.5in]{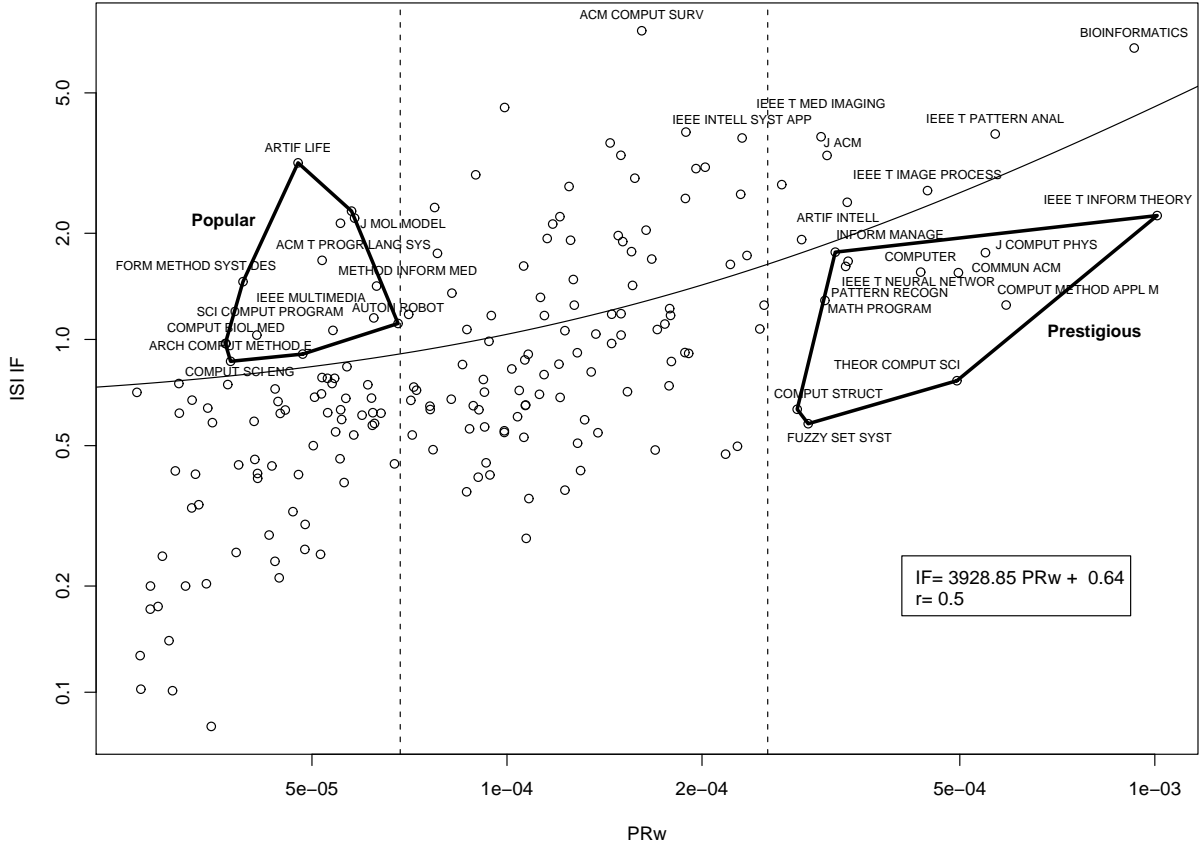}
\end{center}
\caption{\label{CS-outliers} Popular and Prestigious Journals in Computer Science.}
\end{figure}

\begin{figure}[p]
\begin{center}
\includegraphics[width=6.5in]{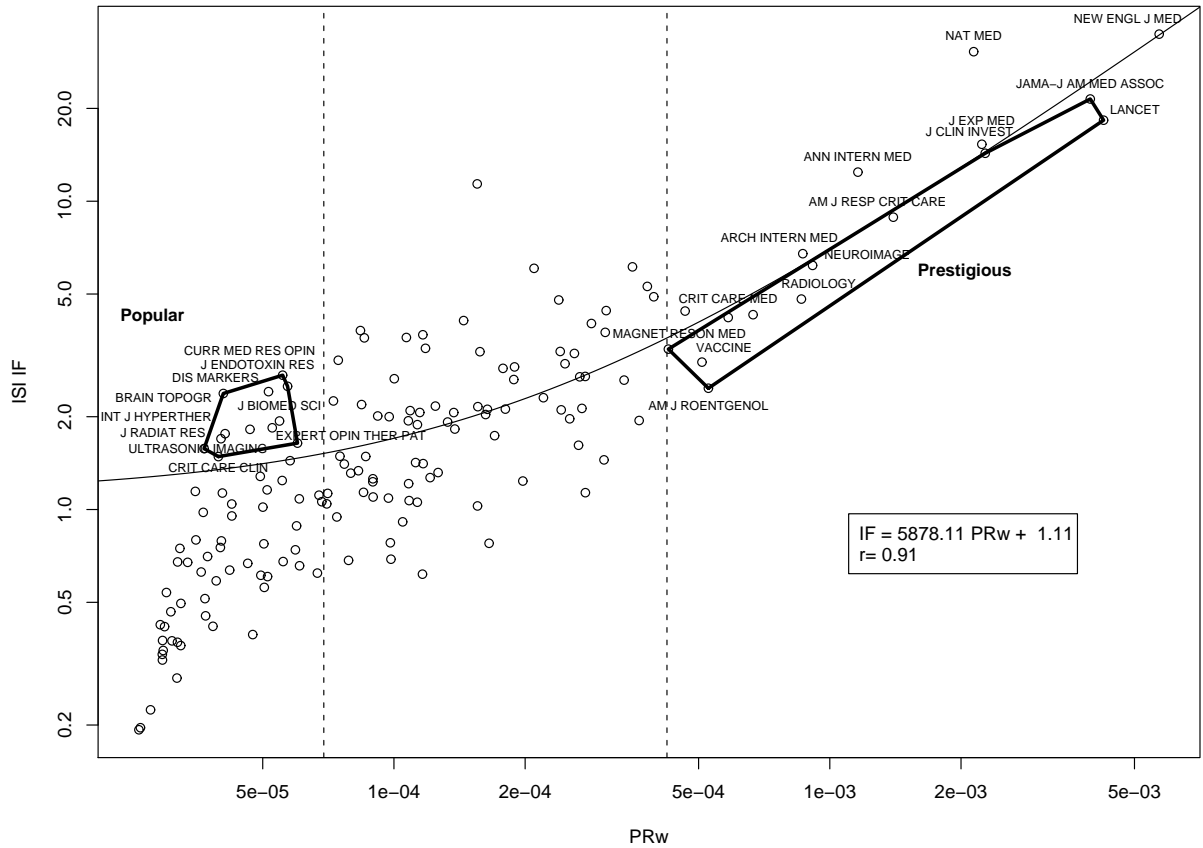}
\end{center}
\caption{\label{med-outliers} Popular and Prestigious Journals in Medicine.}
\end{figure}

Again, a similar analysis was performed for the Computer Science and Medicine subsets of the Journal Citation Network. The Popular journals for Computer Science listed in Table \ref{outliersCS} and shown in Fig. \ref{CS-outliers} contains a number of methodological journals such as Science of Computer Programming and Formal Methods in System Design which may be frequently cited as background material.  It is striking, however, to find that the  journal Artificial Life comes out as the most Popular one; this journal is often cited but not classified as Prestigious.  At the opposite end of the spectrum is the journal IEEE Transactions on Information Theory which is classified as Prestigious, indicating that it is appreciated by domain experts, but it lags in citation counts. When comparing the Popular and Prestigious Medicine journals, shown in Table \ref{outliersmed} and Fig. \ref{med-outliers}, the most striking position is held by Lancet which is assigned to the set of Prestigious journals but whose IF is significantly lower than expected.  Since we had already found that the ISI IF and the Weighted PageRank are most strongly correlated for Medicine, we are not surprised to not find a particularly salient pattern when comparing its Popular and Prestigious sets of journals.\\

\begin{table}[p]
\begin{footnotesize}
\begin{center}
\begin{tabular}{c|lccc||lccc}
	\multicolumn{5}{c}{Popular: ISI IF $\uparrow$, $\mbox{PR}_ w < 40$\%-tile}		&	\multicolumn{4}{c}{Prestigious: ISI IF $\downarrow$, $\mbox{PR}_ w > 90$\%-tile}	\\\hline
   & Journal title					& ISI IF	& PR$_w \times 10^5$	& IF$_\Delta$	& Journal title				& ISI IF	& PR$_ w \times 10^4$	& IF$_\Delta$	\\\hline
1 & ARTIF LIFE 				& 3.17 	& 4.76	& 2.34 					& IEEE T INFORM THEORY	& 2.25	& 10.08 				& $-$2.36 \\
2 & INT J HIGH PERFORM C 		& 2.31 	& 5.76	& 1.44 					& THEOR COMPUT SCI		& 0.76	& 4.95				& $-$1.82 \\
3 & NETWORK$-$COMP NEURAL 	& 2.21 	& 5.82	& 1.34 					& COMPUT METHOD APPL M	& 1.25	& 5.90				& $-$1.71 \\
4 &  J MOL MODEL				& 2.14	& 5.53	& 1.28 					& FUZZY SET SYST			& 0.58	& 2.92				& $-$1.21 \\
5 & ACM T PROGR LANG SYS		& 1.68	& 5.18	& 0.83 					& COMPUT STRUCT		& 0.63	& 2.81				& $-$1.11 \\
6 & FORM METHOD SYST DES	& 1.46	& 3.91	& 0.66 					& COMMUN ACM			& 1.55	& 4.98				& $-$1.05 \\
7 & METHOD INFORM MED		& 1.42	& 6.29	& 0.53 					& J COMPUT PHYS			& 1.76	& 5.48				& $-$1.03 \\
8 & IEEE MULTIMEDIA			& 1.15	& 6.23	& 0.26 					& COMPUTER				& 1.55	& 4.35				& $-$0.80 \\
9 & NEW GENERAT COMPUT		& 1.03	& 4.11	& 0.23 					& MATH PROGRAM			& 1.29	& 3.10				& $-$0.57 \\
10& SCI COMPUT PROGRAM		& 1.06	& 5.39	& 0.21 					& PATTERN RECOGN		& 1.61	& 3.33				& $-$0.34 \\\hline
\end{tabular}
\caption{\label{outliersCS} The top-ranked Popular and Prestigious journals in Computer Science.}
\end{center}
\end{footnotesize}
\end{table}

\begin{table}[p]
\begin{footnotesize}
\begin{center}
\begin{tabular}{c|lccc||lccc}
	\multicolumn{5}{c}{Popular: ISI IF $\uparrow$, $\mbox{PR}_ w < 40$\%-tile}		&	\multicolumn{4}{c}{Prestigious: ISI IF $\downarrow$, $\mbox{PR}_ w > 90$\%-tile}	\\\hline
   & Journal title				& ISI IF	& PR$_w \times 10^5$	& IF$_\Delta$		& Journal title				& ISI IF	& PR$_ w \times 10^3$	& IF$_\Delta$	\\\hline
1 & CURR MED RES OPIN 	& 2.73	& 5.55 & 1.29							& LANCET				& 18.32	& 4.25	& $-$7.76 \\
2 &  J ENDOTOXIN RES		& 2.51	& 5.70 & 1.06							& JAMA					& 21.46	& 3.96	& $-$2.93 \\
3 & DIS MARKERS			& 2.38	& 4.06 & 1.03							& AM J ROENTGENOL		& 2.47	& 0.53	& $-$1.74 \\
4 & ANTISENSE NUCLEIC A	& 2.41	& 5.15 & 0.99							& RADIOLOGY				& 4.82	& 0.86	& $-$1.36 \\
5 & J BIOMED SCI			& 1.94	& 5.46 & 0.50							& VACCINE				& 3.01	& 0.51	& $-$1.10 \\
6 & BRAIN TOPOGR			& 1.82	& 4.67 & 0.43							& INT J RADIAT ONCOL		& 4.29	& 0.67	& $-$0.75 \\
7 & CANCER BIOTHER RADIO & 1.84	& 5.26 & 0.42							& AM J RESP CRIT CARE	& 8.88	& 1.40	& $-$0.46 \\
8 & INT J HYPERTHER		& 1.76	& 4.10 & 0.41							& CRIT CARE MED			& 4.20	& 0.58	& $-$0.36 \\
9 & J RADIAT RES			& 1.70	& 4.01 & 0.35							& MAGNET RESON MED		& 3.31	& 0.43	& $-$0.31 \\
10 & ULTRASONIC IMAGING	& 1.58	& 3.67 & 0.25							& NEUROIMAGE			& 6.19	& 0.91	& $-$0.29 \\\hline
\hline
\end{tabular}
\caption{\label{outliersmed} The top-ranked Popular and Prestigious journals in Medicine.}
\end{center}
\end{footnotesize}
\end{table}

\section{Conclusion}

The distinction between popularity and prestige that is prevalent in all areas of social life has yet to find its way into the assessment of scholarship.  There, the ISI Impact Factor rules as the prime indicator or journal status. The ISI IF for a given journal is based on the number of citations it receives, and ignores the prestige of the citing journals.  Therefore, it is an indicator of journal status that favors popularity over prestige.\\

In this paper, we have added new insights to the ongoing discussions regarding the suitability of the ISI IF as the sole metric of journal status.  The outcome of what is becoming a global discussion can have a fundamental impact on scholarly communication and assessment  \cite{impact:weingart2005}, as the ISI IF metric also lies at the basis of the assessment of the status of scholars, research departments, universities and countries.  In this paper, we found that, while the journal status metric that we obtained by computing Weighted PageRank for all journals in a Journal Citation Network strongly overlapped with the ISI IF, it also revealed significant and meaningful discrepancies.  PageRank is a metric known to take the prestige factor of status into account.  It has provided the foundation for a revolution in Web searching, and it has since successfully been applied to obtain rankings of nodes in a wide variety of networks.  We find the mere fact that the widely used PageRank metric differs in a meaningful manner from the ISI IF a reason to seriously contemplate the use of a variety of journal status metrics instead of just one.  Whether or not a PageRank inspired metric will be added to the status assessment arsenal, it will {\it de facto} change our perception of status as it will be the manner in which scholarly search results will be ranked by Google, Google Scholar and its competitors \cite{tale:smith1999,result:thelwall2001}.\\

To further underline, as many of our colleagues have done before us, that the ISI IF is not the Oracle, but just one of many possible measures of status, we have introduced a ranking of journals according to a product of the ISI IF and the Weighted PageRank. The intuitive and simplistic definition of the Y-factor rankings may not be scientifically convincing, still the authors were more than slightly intrigued to find that the top scoring journals according to this ranking principle rather closely matched their personal perception of importance.

\section*{Acknowledgements} We thank Joan Smith at the Department of Computer Science at Old Dominion University for her assistance in producing the initial raw data on which parts of this analysis are based.

\bibliographystyle{plain}
\bibliography{bollen06scimet}

\end{document}